\documentclass[11pt,a4paper,onecolumn,reqno]{amsart}
\usepackage[a4paper, margin=2.3cm, bmargin=3cm]{geometry}

\usepackage{graphicx,stfloats} \usepackage{color}
\usepackage{amsmath}
\usepackage{amsaddr}
\usepackage{bm, soul}
\usepackage{mathtools}
\usepackage[colorinlistoftodos,prependcaption,textsize=tiny]{todonotes}
\usepackage{braket}
\usepackage{hyperref}
\usepackage[square,comma,sort&compress, numbers]{natbib}
\usepackage{enumerate}
\usepackage[version=4]{mhchem}

\usepackage{amssymb}
\usepackage{adjustbox}
\usepackage{graphicx,stfloats}
\usepackage{color}
\usepackage{float}
\usepackage[numbers]{natbib}
\usepackage[toc,page]{appendix}
\usepackage{color,soul}
\usepackage{caption}
\usepackage{subcaption}
\usepackage{lipsum}
\usepackage{braket}
\usepackage{siunitx}[=v2]
\usepackage{mathabx}
\usepackage{wasysym}
\usepackage[alwaysadjust]{paralist}

\usepackage{placeins}

\newcommand\rev[1]{\textcolor{black}{#1}}

\renewcommand{\st}[1]{}

\usepackage{siunitx}
 
\usepackage{multirow}

\usepackage{etoolbox}
\patchcmd{\pprintMaketitle}
  {\hrule\vskip12pt}
 {\hrule\vskip12pt\ifvoid\extrainfobox\else\unvbox\extrainfobox\par\vskip12pt\fi}
 {}{}

\newsavebox\extrainfobox

\usepackage{caption}
\title{Dynamic stabilization of a hydrogen premixed flame in a narrow channel}

\author{{\large Faizan Habib Vance$^{a,*}$, Arne Scholtissek$^{a}$, Philip de Goey$^{b}$, Jeroen van Oijen$^{b}$, \\ Christian Hasse$^{a}$}\\[10pt]
}
\email{vance@stfs.tu-darmstadt.de} 
\address[]{$^1$Technical University of Darmstadt, Department of Mechanical Engineering, Simulation of reactive Thermo-Fluid Systems, Otto-Berndt-Stra{\ss}e 2, 64287 Darmstadt, Germany\\
$^2$Mechanical Engineering, Eindhoven University of Technology, Eindhoven,The Netherlands
}

\begin{document}
\pagestyle{plain}

\maketitle

\begin{abstract} 
Combustion of hydrogen can help in reducing carbon-based emissions but it also poses unique challenges related to the high flame speed and Lewis number effects of the hydrogen flame. When operated with conventional burners, a hydrogen flame can flashback at higher volumetric flow rates than a methane flame due to the difference in stabilization mechanisms of the two fuels. Due to these differences, conventional burners cannot offer similar operational ranges for hydrogen than that for hydrocarbon flames. An exploration into the unique stabilization behaviour of hydrogen flames is required which could help in envisioning non-conventional burner concepts for keeping hydrogen flames stable. Stability conditions, which describe the kinematics of premixed flames with spatially and temporally changing flow parameters, are crucial for such an exploration. Stability conditions are usually hypothesized for stable flames, where a flame upon perturbation is assumed to return to its original position. Alternatively, in the case of flashback/blow-off, it refers to a flame moving upstream of the burner or being convected out of the domain. However, it is also of interest to understand how and why a flame could move to a new location when the velocity and strain fields are varying with time and space at the original and the new location. In this paper, we investigate the flame stabilization by 1) observing the hydrogen flame's upstream movement in a multi-slit configuration when a geometrical change is made, and 2) changing strain and velocity fields in a dynamic and periodic manner using numerical tools such that the unique behaviour of a hydrogen flame can be captured. We vary the location of high flow strain periodically in a channel by manipulating the boundary condition along a wall. It is found that a hydrogen flame follows this point in a periodic manner, also propagating against the inflow which is considerably faster than its unstretched burning velocity. Spatial and temporal stability conditions, that explain the mechanism behind the flame's movement from its original position to a new position, are analyzed from the simulation data, advancing our knowledge on the flame movement in an unsteady setting and providing important insights into the stabilization mechanism of hydrogen flames. 
\end{abstract}


\section{Introduction} \addvspace{10pt}

\begin{figure}[!h]
\centering
\includegraphics[width=170pt]{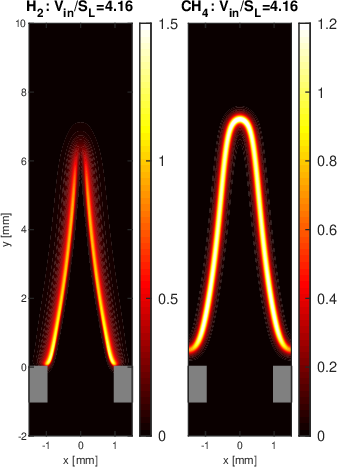}
\caption{Scaled fuel consumption rate for $\ce{H2}$-air and $\ce{CH4}$-air flames stabilized on a multi-slit burner with slit width $W=2 \, \si{\milli\meter}$, distance between plates $D=1 \, \si{\milli\meter}$ and plate thickness of $1 \, \si{\milli\meter}$ at $\phi=0.7$ for both cases.}
\label{fig:H2CH4}
\end{figure}

\begin{figure*}[!h]
\centering
\includegraphics[width=300pt]{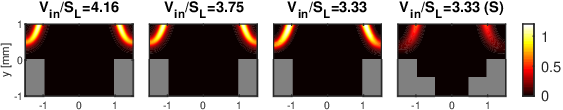}
\includegraphics[width=300pt]{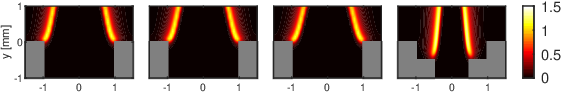}
\includegraphics[width=300pt]{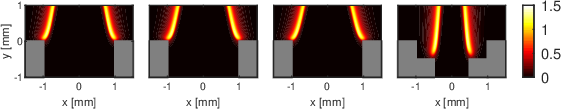}

\caption{Flame base behaviour with changing $V_{in}/S_L$ and addition of a step for $\ce{CH4}$ (DRM19) (top), $\ce{H2}$ (DRM19) (center) and $\ce{H2}$ (Konnov) flames.}
\label{fig:base}
\end{figure*}

The reliable stabilization of a flame on a burner for all of its operating points is essential for a reliable and efficient combustion process. Conventional flame stabilization devices rely on the general idea that a low-speed region created behind a solid object with sharp corners can assist the stabilization of a flame. How exactly a flame anchors depends on the heat transfer with the solid material of the burner, local flow strain rates, the ability of the flame to curve itself and most importantly, in the context of hydrogen enriched flames, the effective Lewis number of the reactants. Due to the differences in thermal and mass diffusivities, fuels can exhibit different responses to flame stretch~\cite{Law1,Poinsotb}. Lean hydrogen flames burn stronger in the presence of positive stretch while the flame speed of fuels with $Le \geq 1$ decreases with the positive stretch rate. The presence of high strain regions near the burner edges can therefore attract the hydrogen flame and provide an anchoring location with a higher speed similar to results found in Refs.~\cite{vance2,cimenez,Dan1}. Ultimately, flame stabilization always results from the kinematic balance between the flow and the flame displacement speed~\cite{vancesc,Poinsotb,state} 

\begin{equation}
    S_{D,Y} = \frac{\rho}{\rho_u} \Big(\mathbf{V_f}-\mathbf{v}\Big) \cdot \mathbf{n}\,, \label{sd}
\end{equation}

\noindent where $\mathbf{v}$ is the local gas velocity, $\mathbf{n}$ is the flame normal vector computed for a scalar field of the reaction progress variable $Y$, $S_{D,Y}$ is the density weighted local flame displacement speed at a location corresponding to a $Y$ iso-level and $\mathbf{V_f}$ is the absolute speed of the flame front relative to the laboratory frame. In order to illustrate the unique stabilization behaviour of the hydrogen flame in contrast to methane flames, a typical hydrogen and a methane flame stabilized on a multi-slit burner are shown in Fig~\ref{fig:H2CH4}. The geometrical setup is based on the recent study by Vance et al.~\cite{vancefb} with slit width $W=2 \, \si{\milli\meter}$, distance between plates $D=1 \, \si{\milli\meter}$ and plate thickness $t=1 \, \si{\milli\meter}$ at $\phi=0.7$ for both flames. The burner temperature is kept constant at $700 \, \si{\kelvin}$ for both $\ce{H2}$ and $\ce{CH4}$ flames. Pressure is kept at 1 atm. Chemistry of $\ce{H2}$-air is modelled using the Konnov mechanism~\cite{Kon} and of $\ce{CH4}$-air is modelled using the DRM19 mechanism~\cite{DRM}. The contours of the fuel source term, which are scaled with the maximum value of the corresponding reference 1D flat flame ($\widehat{\omega_{F}}$), are shown in Fig.~\ref{fig:H2CH4} at the same ratio of inlet flow velocity and the burning velocity ($V_{in}/S_L$). \rev{We tried to find stable flames for both fuels at the same burning velocity and inlet flow velocity but found that it is quite difficult to find stable solutions at these conditions without changing the burner geometry. Instead, we decided to keep the ratio $V_{in}/S_L$ to be constant as this ratio is also relevant to the flame stabilization process.} It can be observed that $\ce{CH4}$ has a slightly longer flame than $\ce{H2}$ and burns slightly stronger at the flame tip. The flame tip does not burn for $\ce{H2}$ while the flame base burns strongly (50 \% higher than the reference 1D flat flame). The $\ce{H2}$ flame is closely attached to the burner corners and the $\ce{CH4}$ flame is lifted from the base with a stand-off distance of almost 0.5 mm. \rev{A comparison of hydrogen and methane flames simulated with the DRM19 mechanism at the same conditions as in Fig.~\ref{fig:H2CH4} is given in the supplementary materials.} In this study our focus is on the flame base and as such, the scaled fuel consumption term is shown in Fig.~\ref{fig:base} with decreasing inlet velocities for both fuels. \rev{The conditions at which the methane and the hydrogen flames were simulated are given in Tab.~\ref{table:multi-conditions}.} The major observations from Fig.~\ref{fig:base} are summarized below:

\begin{table}[]
\caption{\rev{Conditions for the methane and the hydrogen flames shown in Fig.~\ref{fig:base}.}}
\centering
\begin{tabular}{lllll}
\hline
Fuel       & $V_{in}$ {[}m/s{]} & $S_L$ {[}m/s{]} & $V_{in}/S_L$ & $\phi$ \\
\hline
$\ce{CH4}$ & 0.83               & 0.2             & 4.16         & 0.7    \\
$\ce{CH4}$ & 0.75               & 0.2             & 3.75         & 0.7    \\
$\ce{CH4}$ & 0.66               & 0.2             & 3.33         & 0.7    \\
$\ce{H2}$  & 5                  & 1.2             & 4.16         & 0.7    \\
$\ce{H2}$  & 4.5                & 1.2             & 3.75         & 0.7    \\
$\ce{H2}$  & 4                  & 1.2             & 3.33         & 0.7   
\end{tabular}
\label{table:multi-conditions}
\end{table}

\begin{figure}

\centering
\includegraphics[width=200pt]{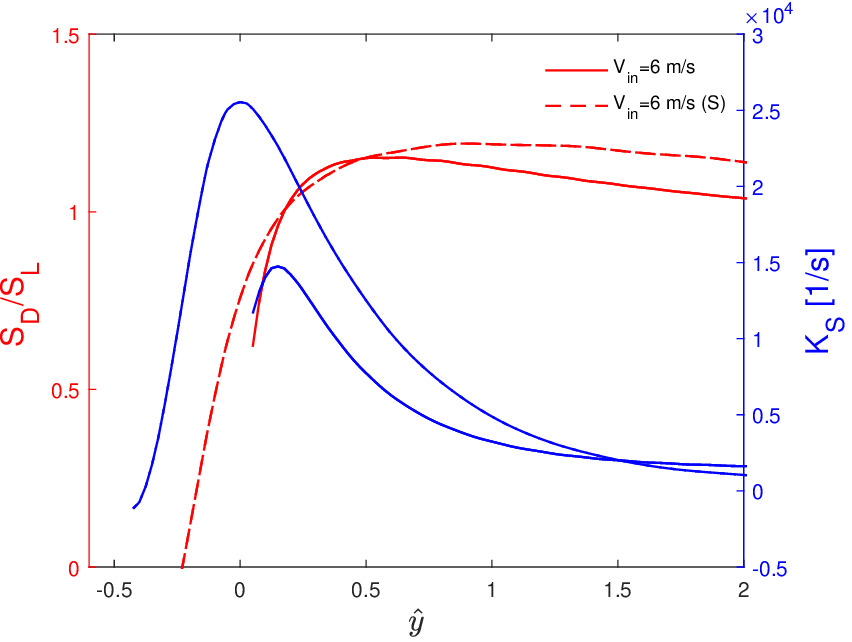}
\caption{Variation of flame displacement speed $S_D$ and stretch rate due to flow strain $K_S$ for $\ce{H2}$-air flame at $V_{in}/S_L=3.33$ with and without an extra step at iso-level of progress variable where the maximum heat release rate occurred.}
\label{fig:H2MBK}
\end{figure}

\begin{itemize}
    \item \rev{Hydrogen flames simulated using the DRM19 and the Konnov mechanisms show identical behaviour.}
    \item It can be observed that the anchoring location of both flames does not change in an observable manner for the three velocity changes. The hydrogen flame remains anchored to the burner corners while the methane flame is lifted about 0.5 mm from the top of the burner surface.
    \item With decreasing inlet velocity, it can be observed that the $\ce{H2}$ flame always anchors near the location of high strain and high velocity, i.e. near a sharp corner.
    \item The $\ce{CH4}$ flame, on the other hand, anchors slightly downstream of the burner in the thermal boundary layer of the top burner surface. In this region the flame exhibits a lower displacement speed due to heat loss and strain rate (which is, lower than that at the burner corner). This matches the lower flow speed in this region and allows for the flame to remain anchored at the same location even with decreasing inlet velocity.
    \item Next, if we insert a geometrical step as shown for the cases with '(S)', keeping the same inlet velocity, we observe an interesting behaviour of the two flames. The $\ce{CH4}$ flame stays anchored at the same position but burns weaker than the flame without the extra step. This is caused by the increase in the heat loss to the higher burner surface area. The $\ce{H2}$ flame on the other hand, moves to the new location of higher strain and higher velocity.
    \item The contribution from strain towards the total stretch can be calculated as~\cite{Poinsotb} $K_S= \nabla_t \cdot \mathbf{v}$, where $\nabla_t$ is the tangential component of the $\nabla$ operator and $\mathbf{v}$ is the local flow velocity vector. In Fig.~\ref{fig:H2MBK}, $K_S$ is plotted at the iso-level of progress variable $Y$ where the maximum heat release rate occurred. It is found that the maximum value of strain near the flame base increases with the addition of a geometrical step. The ratio $S_D/S_L$ (plotted at the same respective iso-levels) shows that the flame speed slightly decreases at the location of maximum stretch  and this is caused by the higher heat loss to higher surface area of the modified burner.
    \item This movement of the $\ce{H2}$ flame is quite unique and is based on similar observation in Refs.~\cite{vancenrz,yuri,Dan1,cimenez} where addition of $\ce{H2}$ into $\ce{CH4}$-air mixtures resulted in flame moving further upstream. However, here it shows clearly that the $\ce{H2}$ flame has a unique attribute to propagate upstream to a more favorable anchoring location with higher strain and flow velocity.
\end{itemize}

\begin{figure}
  \centering
{\includegraphics[height=85pt]{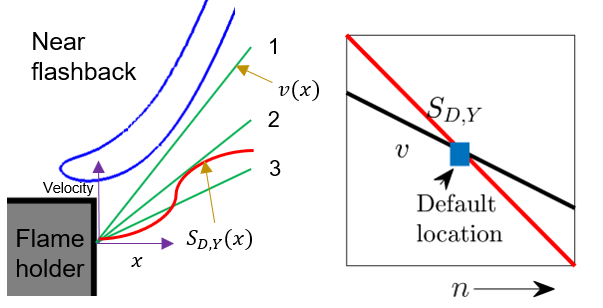}}
      \caption{Illustration of stability conditions according to the critical velocity gradient theory (left) and dynamic stability criterion (right) as discussed in the text.}
      \label{fig:gcillus}
\end{figure}

In order to devise a numerical setup in which we can control the flame movement in an isolated manner, a brief overview of stability conditions from the literature is presented. The pioneering study on the understanding of stability conditions was done by Lewis and von Elbe~\cite{Elbe1} for flames stabilized on a circular tube burner with a critical velocity gradient condition. A graphical illustration of the critical velocity gradient concept is shown schematically in Fig.~\ref{fig:gcillus} for a flame near flashback. Different gas velocity profiles at the burner side wall are assumed as locally linear and are indicated by green curves. The red line describes the local flame speed $S_{D,Y}$ and the flame zone is visualized with a blue line. The $x$ coordinate is the horizontal coordinate and $v(x)$ is the vertical velocity. With velocity profile 1, the gas velocity is greater than $S_{D,Y}$ at all locations and the flow can push the flame away from the burner. For velocity profile 2, $v(x)$ and $S_{D,Y}$ are equal at a certain point and the flame  stabilizes there. The velocity gradient of curve 2 is the critical velocity gradient for flame flashback. For velocity profile 3, gas velocity $v(x)$ is less than $S_{D,Y}$. This causes the flame to move upstream resulting in flame flashback. The critical velocity gradient $\partial v/\partial x$ can thus be related to $S_{D,Y}$ at a distance equal to flame thickness $\delta_F$ in the x-direction from the burner wall for stable flames as

\begin{equation}
    \Bigg\lvert \frac{\partial v}{\partial x}\Bigg\lvert \geq \frac{ S_{D,Y}}{ \delta_F}.
\end{equation} 

\begin{figure}
  \centering
{\includegraphics[height=110pt]{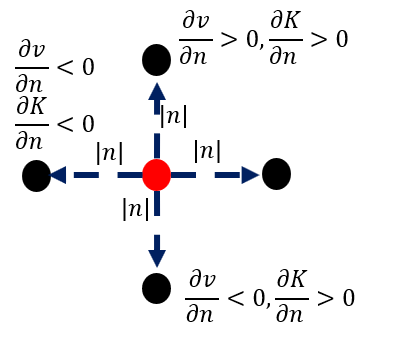}}
      \caption{Illustration of an anchoring point (or the default flame location) $\color{red} \CIRCLE$ with four possible points $\color{black} \CIRCLE$ where the flame could move in an adiabatic situation.}
      \label{fig:ancillus}
\vspace{-8pt}
\end{figure}

\noindent The above condition describes a critical value (when the left and the right hand sides are the same) for the flame to stabilize on top of the burner, but it does not describe whether a flame will move to its original position when perturbed. For this purpose, another stability condition, which has been discussed for premixed flames, is the dynamic stability condition~\cite{Kedia2,kawa1}. This condition states that if the flame is perturbed such that the gradient of $S_{D,Y}$ in the flame normal direction is greater than the gradient of velocity, the flame will return to its original position (right side of Fig.~\ref{fig:gcillus}). This condition can be mathematically written as 
\begin{equation}
     \Bigg\lvert \frac{\partial v}{\partial n}\Bigg\lvert < \Bigg\lvert \frac{\partial S_{D,Y}}{\partial n}\Bigg\lvert.
\end{equation}

\noindent This condition was further analyzed in Ref.~\cite{Kedia2} by decomposing the displacement gradient into a Markstein length~\cite{mat2} and the gradient of stretch $K$ in the flame normal direction as

\begin{equation}
    \frac{\partial S_{D,Y}}{\partial n}=\frac{d S_{D,Y}}{d K} \frac{\partial K}{\partial n}. \label{eq:dynstab}
\end{equation}

\noindent The above equation incorporates the effect of the fuel Lewis number in the form of a sensitivity coefficient, the Markstein length $\mathcal{L}_M = d S_{D,Y}/dK$, which can be estimated using canonical 1D configurations and contains the essential information to describe the flame returning to its original position upon perturbation in a quasi-steady state manner. For an adiabatic environment, the flame movement can depend strongly on the stretch and velocity fields ahead and behind of the flame as a function of space and time. Such a situation is illustrated in Fig.~\ref{fig:ancillus} where the flame anchoring/default location, upon perturbation, can move towards one of the four possible regions characterized by velocity and stretch. For $\mathrm{Le} < 1$ flames, the flame can move towards a region of higher flow speed if its flame speed $S_{D,Y}$ also increases with increasing stretch rate. Conversely, for flame movement towards the lower speed region, its flame speed should decrease with decreasing stretch rate (also applicable for $\mathrm{Le}<1$ flames). Furthermore, with varying time, stretch/strain and velocity at the default flame location could change requiring the flame to move to a new position. For $\mathrm{Le} \geq 1$ flames, the flame could move towards a region of high/low strain and low/high velocity resulting in an decrease/increase in $S_{D,Y}$ to find a new stabilization location. 

There have been numerous studies on flame stabilization of premixed hydrogen flames~\cite{KLUKAS202032547,GOLOVASTOV20212783,ZHANG20201072,ELSHIMY202014979,WU202133601,ZOU202020436,MARRAGOU202219275,en14071977,LEE20152602,JIMENEZ201512541,REICHEL20174518,ENDRES2018151,GOLDMANN2022111927, WAN2020358, CHEN20096558,SHI2016394,doi:10.1080/13647830.2013.792393,ZHANG2017170,ZHANG20211955,GERLINGER2003247}. However, the literature discussing flame movement in a semi-controlled manner is rather scarce. In order to understand the dynamic stabilization of premixed hydrogen flames we study their movement inside a rectangular channel. The movement of the flame is controlled by the use of a boundary condition employing periodic movement of a high shear stress ($\propto$ flow strain) point at a wall. Notably, the hydrogen flame is also shown to follow the point of high shear stress when propagating against the inflow, which is considerably faster than its unstretched burning velocity. The flame in a channel has been widely studied in the past with focus on meso-scale combustion~\cite{maruta}, symmetrical and asymmetrical flames~\cite{kurd6}, flame acoustics~\cite{spain}, effect of gas compressibility~\cite{kurd}, and flame propagation behaviour~\cite{ronney,vervisch}. In this study, we add another example where the flame in a channel configuration yields important insights on flame physics, owing to its simplicity yet covering also complex effects of dynamic flame stabilization. 

The objective of this study is to elucidate in a clear manner, the ability of a hydrogen premixed flame to move towards the preferred anchoring location of high strain and high velocity. In order to show this behaviour, the flame in a channel configuration is used with a modified boundary condition to generate a region of high strain and high velocity. In this way, the movement of the hydrogen flame observed in Fig.~\ref{fig:base} can be studied in isolation from the heat loss effect. This study is organized is as follows: In Sec.~2 we introduce the numerical employed in this study and in Sec.~3, results for dynamically stabilized hydrogen flames are discussed. Section 4 discusses the dynamic flame stabilization process by analyzing the reacting and non-reacting solutions. Section 5 concludes this study.

\section{Numerical model and initial conditions}  \addvspace{10pt}

\begin{figure*}[h]
\centering
\includegraphics[width=340pt]{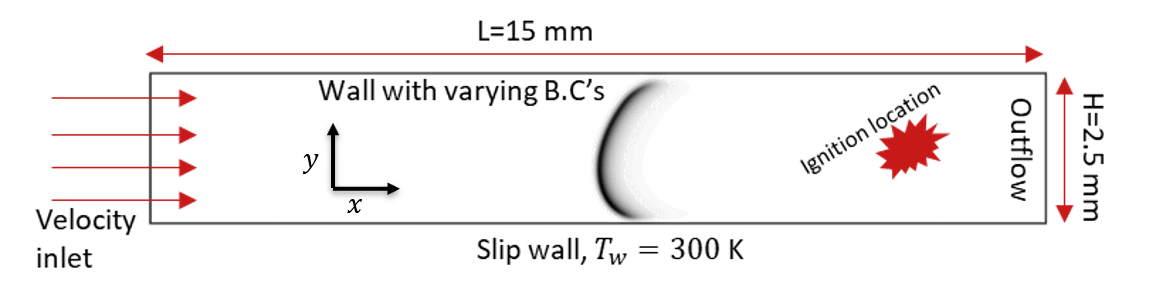}
\vspace{-3pt}
\caption{Computational domain of the flame in a channel configuration.}
\label{fig:model1}
\end{figure*}

The computational model used in this study consists of a rectangular channel with length $L=15 \, \si{\milli\meter}$ and height $H=2.5 \, \si{\milli\meter}$. A schematic of the domain is given in Fig.~\ref{fig:model1} where the location of flame ignition and an instant during the flame propagation towards the inlet is shown. Premixed fresh gases enter the domain from the left with a uniform inlet velocity and a temperature of $300 \, \si{\kelvin}$. The outlet is modelled with a Neumann type boundary condition implying that there is no change in the field variables in the normal direction. The bottom wall is prescribed as a slip boundary and the temperature is equal to the inlet value. \rev{Thereby, the flame is prevented from anchoring near the bottom wall as the prescribed temperature would lower the flame speed due to heat loss but the flow velocity remains higher than the burning velocity due to the slip condition.} The top wall boundary condition is modified such that artificial anchoring conditions can be applied. This will be discussed in detail in Sec.~2.2. Gravitational and viscous work effects are neglected. In the detailed chemistry simulations, Soret or thermal diffusion effects are modelled by using a reduced model for $\ce{H}$ and $\ce{H2}$ species following~\cite{model,vance2}. Radiation heat loss from the gas to the surrounding environment is also modelled using an optically thin model~\cite{rad}. Soret diffusion and radiation heat losses were found to play no significant qualitative effect on the observations made in this paper and are included here for the sake of completeness of the model. The chemistry of the $\ce{H2}$-air flames is modelled using the Konnov mechanism~\cite{Kon} which contains 15 species and 75 reactions and Li's mechanism~\cite{Li} with 9 species and 21 reactions. Ansys Fluent~\cite{Ansys} is used to solve the unsteady flames using an unsteady coupled solver. \rev{We have used second order upwind schemes for momentum, energy and species equations while second order scheme is used for the pressure equation along with a second-order time integration scheme.} The two-dimensional unsteady reacting flow equations are solved on an equidistant Cartesian grid with a $25 \, \si{\micro\meter}$ global grid resolution. A time step of $10 \, \si{\micro\second}$ is used throughout this study resulting in a Courant number less than 1. \rev{Direct integration of the chemical source term is employed for a high quality solution along with a stiff chemistry solver. }For the unsteady laminar reactive flow with low-Mach formulation, the following equations are solved:

\begin{equation}
\frac{\partial \rho}{\partial t} + \nabla \cdot(\rho \mathbf{v})=0,
\end{equation}
\begin{equation}
\frac{\partial \rho Y_i}{\partial t} +\nabla \cdot (\rho \mathbf{v} Y_i)+ \nabla \cdot \mathbf{J}_i=\omega_i,
\end{equation}
\begin{equation}
\frac{\partial \rho \mathbf{v}}{\partial t} +\nabla \cdot (\rho \mathbf{v} \mathbf{v})=-\nabla p + \nabla \cdot \bar{\bar{\tau}},
\end{equation}
\begin{equation}
\frac{\partial \rho E}{\partial t} +\nabla \cdot ((\rho E+p)\mathbf{v})=\nabla \cdot(\lambda \nabla T) - \nabla \cdot (\Sigma_i h_i \mathbf{J}_i)+ \omega_T + Q_{rad}.
\end{equation}

\noindent In the above equations, the velocity vector, density, the species mass fractions, the species source terms, pressure, temperature, species sensible enthalpy and heat loss due to radiation from burnt gases ($\ce{H2O}$ vapor) are represented by $\mathbf{v}$, $\rho$, $Y_i$, $\omega_i$, $p$, $T$ , $h_i$ and $Q_{rad}$, respectively. The stress tensor, total energy, enthalpy and thermal heat release rate are represented by $\bar{\bar{\tau}}$, $E$, $h$ and $\omega_T$, respectively and modelled as

\begin{equation}
\bar{\bar{\tau}}=\mu [(\nabla \mathbf{v}+ \nabla \mathbf{v^T})-\frac{2}{3} \nabla \cdot \mathbf{v} I],
\end{equation}
\begin{equation}
E=h-\frac{p}{\rho},
\end{equation}
\begin{equation}
h=\sum_{i} Y_i h_i,
\end{equation}
\begin{equation}
h_i=\int_{T_{ref}}^{T} c_{p,i} dT,
\end{equation}
\begin{equation}
\omega_T=\sum_{i} \frac{h_i^0}{M_i} \omega_i.
\end{equation}

 \noindent Here, formation enthalpy and molecular weight of species $i$ are represented by $h_i^0$ and $M_i$, respectively. $\mathbf{J}_i$ represents the diffusion flux and is given by
\begin{equation}
   \mathbf{J}_i= \mathbf{J}_i^F+\mathbf{J}_i^T.
\end{equation}

\noindent where the Fickian diffusion flux due to species gradients is given as: \rev{$\mathbf{J}_i^F=- \rho D_{i} \nabla Y_i$,  with $D_{i}$ Fickian diffusion coefficients for species $i$ calculated from constant Lewis numbers.} The thermal diffusive flux (Soret effect) due to temperature gradients is given as: $\mathbf{J}_i^T=- \rho D_i^T \frac{\nabla T}{T}$, with $D_i^T$ being the thermal diffusion coefficient for species $i$. \rev{Constant Lewis numbers, calculated by simulating one-dimensional flat flames with multi-component transport model using CHEM1D~\cite{chem1d} are used for mixture properties similar to~\cite{vance1,vance2} for faster computation and adequate accuracy. Constant non-unity Lewis number approach has shown to work adequately for flames with strong differential diffusion effects in~\cite{vance2,ConstLE,state,oijen1}. Lewis numbers are calculated based on the local Fickian diffusion flux, $\mathbf{J}_i^F$ using the multi-component model by,}

\begin{equation}
    Le_i=-\frac{\nabla Y_i}{\mathbf{J}_i^F}  \frac{\lambda}{c_p}.
    \label{eq:le1}
\end{equation}

\noindent \rev{Here $Le_{i}$ are the constant Lewis numbers for species $i$ and $\lambda$ and $c_p$ the mixture conductivity and the specific heat capacity. Lewis numbers used for each species for $\ce{CH4}$-air and $\ce{H2}$-air flames are given in the supplementary materials.} The transport properties are calculated based on the following relations~\cite{oijen1,vance2}:

\begin{equation}
{\lambda}= \num{2.58e-5} \, c_p \,  \left(\frac{T}{298} \right)^{0.69} [\si{\watt\per\meter\per\kelvin}],
\end{equation}
\begin{equation}
{\mu}= \num{1.67e-8} \, c_p \,  \left( \frac{T}{298} \right) ^{0.51} [\si{\kilogram\per\meter\per\second}].
\end{equation}
\label{eq:le}

\noindent Here, the mixture viscosity is represented by $\mu$. \rev{For the fitted curves, values of $c_p$ and temperature are required to be entered in the right units for evaluation of mixture viscosity and conductivity.} \rev{We have implemented the mixture conductivity, viscosity and diffusion coefficients of species with the help of user-defined-functions (UDFs) in the Ansys Fluent.}

\subsection{Shear stress specification at the top boundary} 

In order to control the location of the maximum strain at the top wall, the shear stress $\tau$, given by the following equation, can be specified at the top wall 

\begin{equation}
    \tau(x,t)=- \mu \frac{\partial u}{\partial y} \,,
\end{equation}

\noindent where $y$ is the vertical coordinate, $\mu$ is the dynamic viscosity and $u$ is the velocity in the $x$ direction. The shear stress at the wall is prescribed with the following relation


\begin{figure*}[!h]
\centering
\includegraphics[width=320pt]{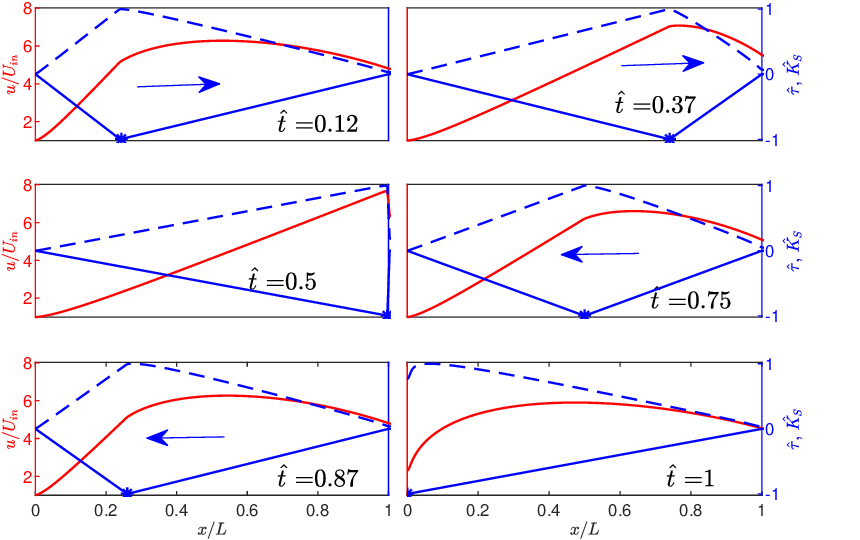}
\vspace{1pt}
\caption{Cold flow result at the top wall for one time period. Results are shown for scaled $x$-velocity (red) and normalized shear stress $\widehat{\tau}$ (solid blue) and strain $\widehat{K_S}$ (dashed blue). The location of the maximum normalized stress is marked with $\Asterisk$. Arrows indicate the direction of movement of $x^*$. }
\label{fig:TauCold}
\end{figure*}

\begin{equation}
     \tau(x,t)=\begin{cases}
    -A \frac{x}{x^*}, & \text{if $x \leq x^*$}.\\
    -A \frac{x-L}{x^*-L}, & \text{if $x > x^*$}.
  \end{cases}
\end{equation}

\noindent Here, $x^*(t)$ is a location where the shear stress has its minimum along the top wall and moves with time. $A$ is the magnitude of the applied shear stress. The point where $\tau$ is minimum, $x^*$, is calculated from the following equations for forward and rearward moving waves

\begin{subequations}
\begin{equation}
    x_i^*=\frac{2Lt}{TP}-(i-1)L, \: \: \: \text{for $i=1,3,5 $},
\end{equation}

\begin{equation}
    x_j^*=-\frac{2Lt}{TP}+(j)L, \: \: \: \text{for $j=2,4,6 $}.
\end{equation}

\end{subequations}

\noindent $x^*$ is chosen from the values of $x_i$ and $x_j$ which lie inside the range $0 \leq x \leq L$. In this study, we only consider three time periods for the $x^*$ to move from the inlet to the outlet and back. The time period $TP$ can be calculated as

\begin{equation}
    TP=\frac{2L}{c},
\end{equation}


\noindent with $c$ being the speed of the wave. \rev{The boundary condition for stress is implemented as a UDF in Ansys Fluent.} A result from an unsteady cold flow simulation with $U_{in}=2 \, \si{\meter\per\second}$ is shown in Fig.~\ref{fig:TauCold} for one time period at different instances of scaled time $\hat{t}=t/TP$. The velocity is scaled with the inlet value while the shear stress is scaled with $A=2$ Pa and the strain $K_S$ is scaled with its local maximum value indicated by hats over symbols. The location of minimum $\tau$ and maximum $K_S$ can be observed to move from the inlet to the outlet. The corresponding change in the velocity profiles can be observed with location of maximum velocity being slightly ahead of $x^*$. With $\hat{t}>0.5 $, the wave reflects back towards the inlet where the location of maximum value trails the $x^*$ point.

\subsection{Initial conditions} 

\begin{figure}[b!]
\centering
\includegraphics[width=\columnwidth]{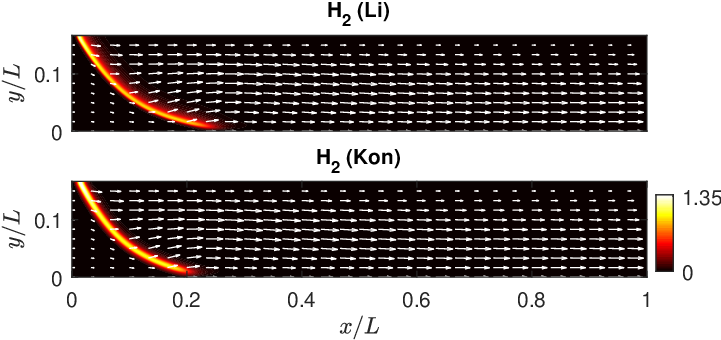}
\caption{Scaled fuel consumption rate $\widehat{\omega}_F$ of the $\ce{H2}$ flames at the same $U_{in}/S_L$ with Li's mechanism (top) and Konnov's mechanism (bottom) with a no-slip condition at the top wall. Velocity vectors are shown in white.}
\label{fig:hrr-IC}
\end{figure}

In order to proceed with the unsteady reactive flow simulations, initial conditions for the hydrogen flame need to be generated such that the flame stabilizes closer to the inlet. A hydrogen flame is simulated at $\phi=0.7$ with the $x$ inlet velocity $U_{in}=2 \, \si{\meter\per\second}$ and burning velocity $S_L=1.2 \, \si{\meter\per\second}$. The boundary condition for the top wall is prescribed as a no-slip and zero-heat flux (adiabatic) condition. This generates artificial flow strain at the node common to the inlet and the top wall, thus allowing the flame to lower/increase its flame speed depending on the fuel Lewis number.


The fuel consumption rate is scaled with the corresponding maximum consumption rate (absolute) of a flat unstretched flame and the scaled quantity, $\widehat{\omega_F}$, is shown in Fig.~\ref{fig:hrr-IC} with superimposed velocity vectors using Li and Konnov mechanisms. The hydrogen flame is found to burn 30-40 \% stronger than the reference adiabatic unstretched flame with both mechanisms. The flames are asymmetrical, stabilize closer to the inlet at the top wall and lose heat at the bottom wall due to the constant temperature boundary condition. Overall, both flames are anchored close to the inlet and show an acceptable comparison with each other.


\section{Results and discussions} \addvspace{10pt}

\begin{figure*}[!h]
\centering
\includegraphics[width=175pt]{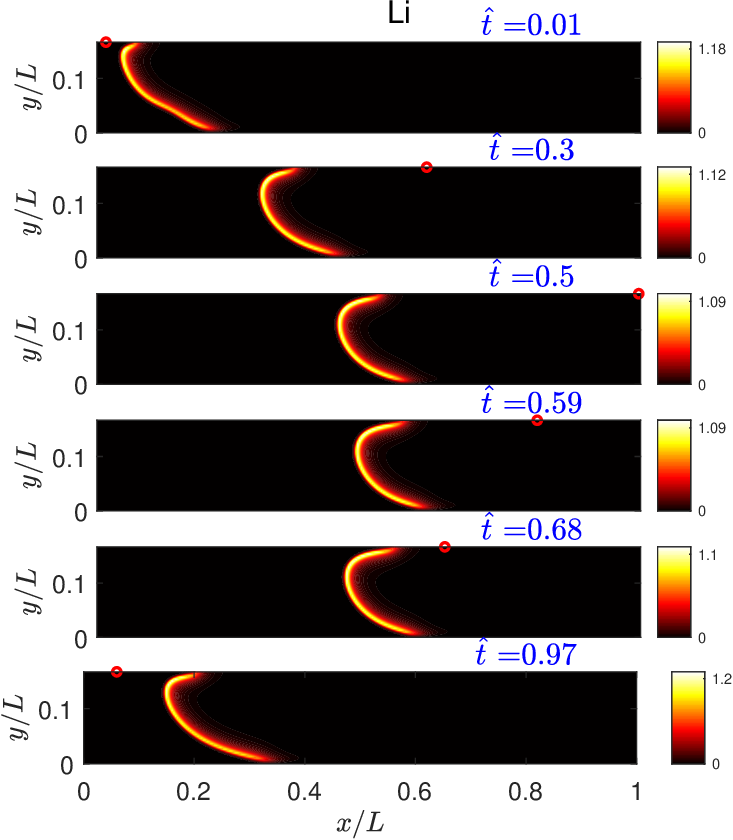}
\includegraphics[width=175pt]{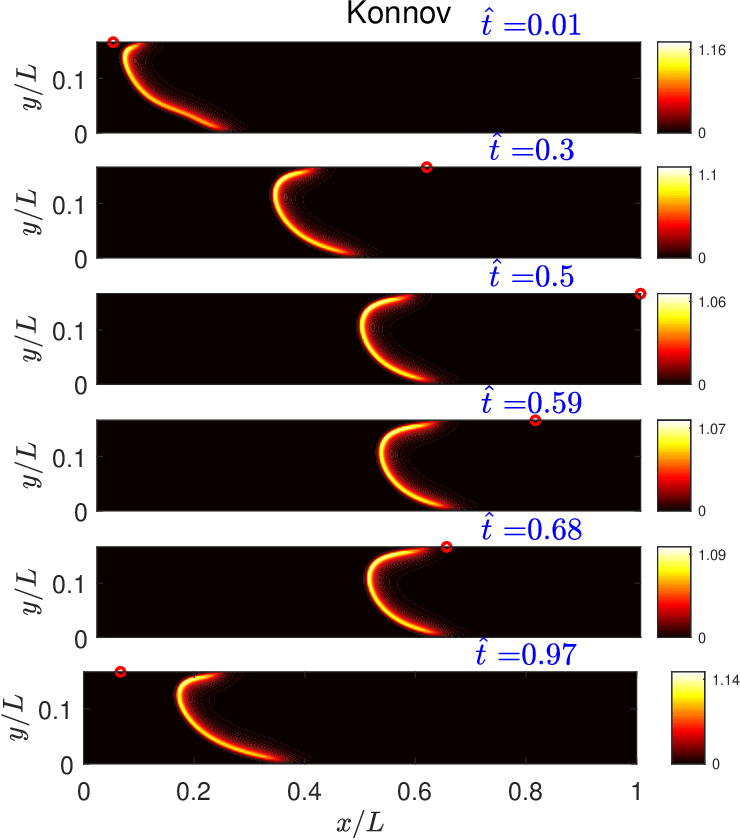}
\caption{Scaled fuel consumption rate $\widehat{\omega}_F$ for Li (left) and Konnov (right) mechanisms at the same time-steps. The location of maximum strain (minimum stress) $x^*$ is visualized with a red circle on the top wall. A video of the time sequences is available in the supplementary material.}
\label{fig:Dyn-TwoFuels}
\end{figure*}

In this section, results from the unsteady simulation are discussed which use the initial conditions together with the top wall boundary condition described in the previous section. First, a comparison is made between the usage of the two mechanisms for the hydrogen flame at the same $U_{in}/S_L$ with the same wave speed of $c=S_L/6$ for the location with minimum shear stress at $x=x^*$. Second, the wave speed $c$ is varied with the Konnov mechanism in order to understand its effect on the periodic flame movement.

\subsection{Hydrogen flames with two different kinetic schemes} 

\begin{table}[]
\caption{\rev{Conditions for the hydrogen flames discussed in this subsection}}
\centering
\begin{tabular}{|l|l|}
\hline
$\phi$     & 0.7           \\
$S_L$      & 1.2 {[}m/s{]} \\
$U_{in}$   & 2 {[}m/s{]}   \\
Pressure   & 1 atm         \\
Re         & 118           \\
Wave speed & $S_L/6$    \\  
\hline
\end{tabular}
\label{tab:channel-conditions}
\end{table}

\begin{figure}[b!]
\centering
\includegraphics[width=200pt]{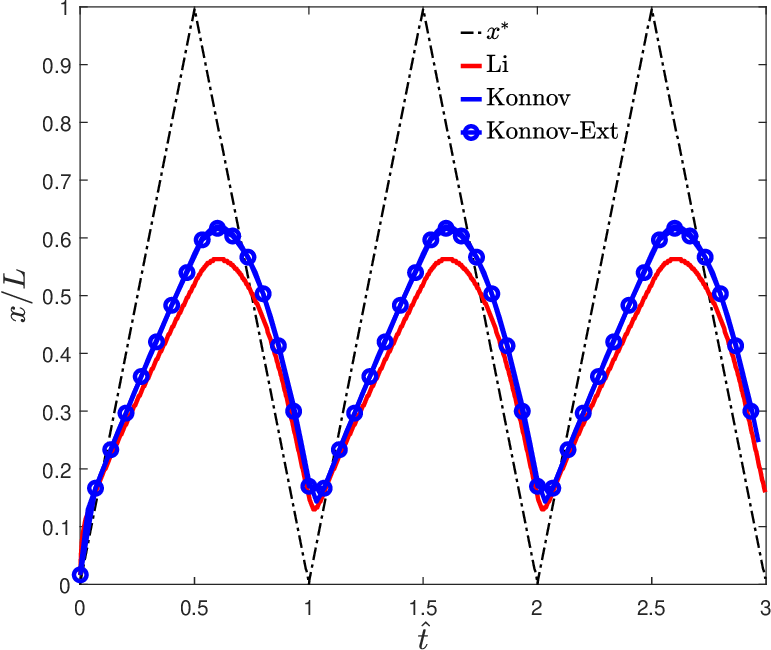}
\caption{Variation of the flame position $x_f$ (solid lines) on the top wall along with location of minimum applied stress (dashed line) for hydrogen flames $c=S_L/6$ using two different kinetic mechanisms. \rev{A validation case with the outlet extended by 5 mm is also presented using the Konnov mechanism.}}
\label{fig:mech}
\end{figure}

Conditions for the hydrogen flames presented in this subsection are summarized in Tab.~\ref{tab:channel-conditions}. \rev{Reynolds number (Re) calculated based on the inlet conditions gives a value of 118. Pressure is kept at 1 atm throughout this study.} Results for the first time period are shown in Fig.~\ref{fig:Dyn-TwoFuels} with the two kinetic schemes at the same $U_{in}/S_L=1.66$ with the same wave speed of $c=S_L/6$. The location of the minimum applied stress $x^*$ is represent by a red circle on the top wall. One time period is completed in 0.15 seconds and $A$ is kept at 2 Pa for hydrogen. The hydrogen flame follows the red circle towards the outlet direction and when it is reflected, the flame follows. A video of the flame moving periodically in a channel is available in the supplementary materials. The hydrogen flame did not move beyond $x/L \approx 0.6$ in the simulation but follows the general direction of the prescribed point. It is further observed that the hydrogen flame can adapt its apparent curvature at different time intervals. The maximum value of $\widehat{\omega_F}$ also changes during the flame movement. Overall, an excellent qualitative comparison is found between the mechanisms during the different flame sequences. A more quantitative comparison is shown in Fig.~\ref{fig:mech} where the flame position at the top wall is tracked for the results using the two mechanisms. The flame position $x_{f}$ is identified by the location of the maximum $\widehat{\omega_F}$ close to the wall at $y/L \approx 0.99$. It can be observed that the flame closely follows the movement of the $x*$ point and moves downstream and upstream in a periodic manner. Results from both mechanisms show good agreement with a slightly under-prediction in the maximum flame location with Li's mechanism. It can be concluded that deviations between the two kinetic mechanisms are minor, and small optimized mechanisms, such as the Li mechanism, can reproduce the flame dynamic movement behaviour sufficiently well. We will use the more detailed Konnov mechanism through the rest of this paper. An analysis of the hydrogen flame presented in his section will follow in Sec. 3.3.

\subsection{\rev{Effect of Neumann condition at the outlet}} 
\rev{
In this subsection, we will briefly show the impact of Neumann condition at the outlet on the flame movement by comparing results from the model shown in Fig.~\ref{fig:model1} with the same model but with an extension of 5 mm at the downstream section. The boundary condition for applied stress is only applied till the original length and stress is put to zero for the extended section of the top wall. Results are shown in Fig.~\ref{fig:mech} using the Konnov mechanism for both the original and the extended model. We can observe that there is an excellent comparison between the two models and we can conclude that for the usage of the Neumann condition at the outlet does not impact the flame dynamics in a major way. }

\subsection{Hydrogen flame at varying wave speed} 

\begin{figure}[b!]
\centering
\includegraphics[width=200pt]{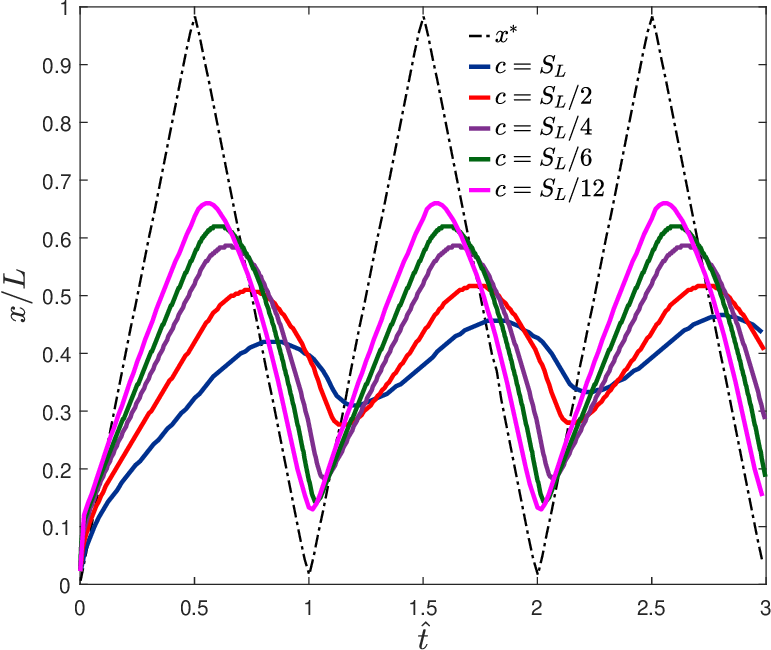}
\caption{Variation of the flame position $x_f$ (solid lines) on the top wall along with location of minimum applied stress (dashed line) for hydrogen flames at $U_{in}=2 \, \si{\meter\per\second}$ and different wave speeds $c$.}
\label{fig:wavespeed}
\end{figure}

In this subsection, the wave speed $c$ is varied from $c=S_L$ to $c=S_L/12$ for the hydrogen flame. This increases the time period from $0.025 \, \si{\second}$ to $0.3 \, \si{\second}$. Three time periods are simulated for each case to verify that the flame is stable in a periodic manner. The flame position $x_{f}$, identified by the location of the maximum $\widehat{\omega_F}$ close to the wall at $y/L \approx 0.99$, is plotted in Fig.~\ref{fig:wavespeed} together with the systematically varied $x^*$. It is observed that for $c=S_L$, the flame initially moves to the middle of the domain and only moves up and downstream within the range of $\pm 0.1$. The peak location also increases with the time period, indicating that a periodic stabilization has not been reached with the wave speed equal to $S_L$. Decreasing the wave speed to $S_L/2$, it can be observed that the flame appears to stabilize dynamically in a periodic manner with the difference between maximum and minimum of $x_f$ increases. With a further decrease in the wave speed, a periodic movement of the flame can clearly be identified and the difference between maximum and minimum $x_f$ is around $0.4$ for $c=S_L/6$ and around $0.45$ for $c=S_L/12$. This dynamic stabilization of the hydrogen flame indicates that it can be stable for an indefinite amount of time depending on the wave speed of the applied motion. In all of the cases presented, the flame does not move beyond $x/L=0.65$ for $c=S_L/12$ and there appears to be a time lag between $x^*$ and $x_{f}$ locations. In the next subsection, we analyse the local flame structure for the case $c=S_L/6$.

\subsection{Flame structure analysis} 

\begin{figure}[b!]
\centering
\includegraphics[width=200pt]{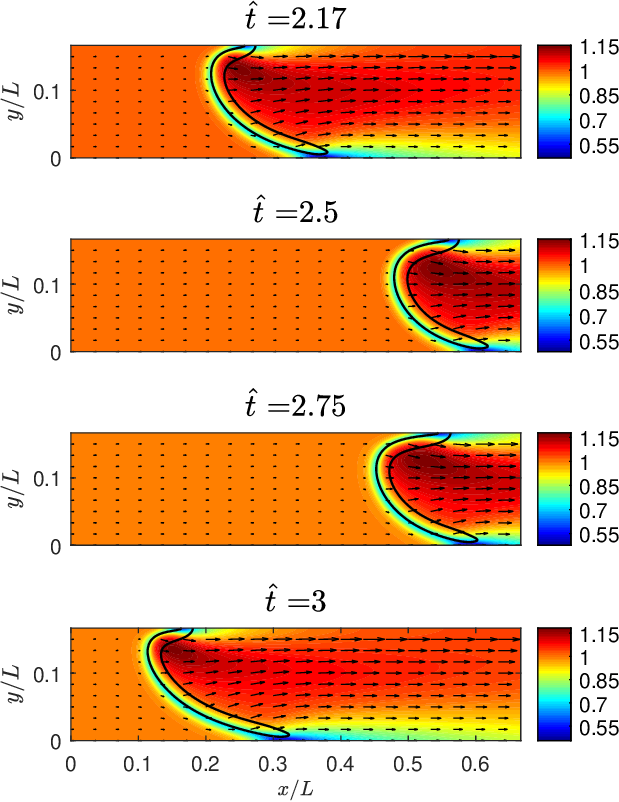}
\vspace{-5pt}
\caption{$Z_{\ce{H}}/Z_{\ce{H},u}$ at $TP=2.17$ with superimposed flow-vectors and an iso-level of 20 \% of maximum fuel consumption rate at. various instances of the third time period.}
\label{fig:ZH}
\end{figure}

In this subsection, local profiles for the $c=S_L/6$ case are analyzed to understand the flame response to the applied shear stress. In order to visualize the preferential diffusion effects, the hydrogen elemental mass fraction $Z_{\ce{H}}/Z_{\ce{H},u}$ is plotted in Fig.~\ref{fig:ZH} at various instances of the third time period. $Z_{\ce{H},u}$ is the hydrogen elemental mass fraction of the unburnt mixture. The flame location is represented with an iso-level of the fuel consumption rate (20 \% of the maximum value in the domain). It shows that the flame is curved, especially near the top wall region. In the immediate vicinity of the top and bottom walls, a leaner mixture is present due to negative curvature similar to flame cusps. Here, due to preferential diffusion, all the fuel diffuses towards the immediate sides of the flame, where we can observe the flame to have a positive curvature and $Z_{\ce{H}}/Z_{\ce{H},u}$ to be around 1.15 for all the time instances.

\begin{figure}[!h]
\centering
\includegraphics[width=\columnwidth]{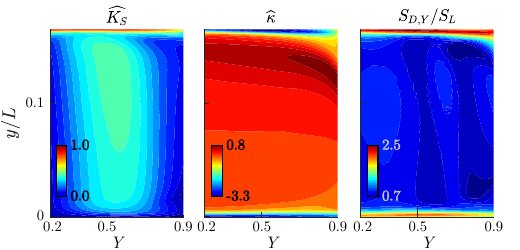}
\caption{Contour plots of the scaled quantities $\widehat{K_S}$, $\widehat{\kappa}$, and $S_{D,Y}/S_L$ as a function of progress variable $Y$ and scaled vertical length $y/L$ at $\hat{t}=2.17$ for $c=S_L/6$.}
\label{fig:Profiles}
\end{figure}

Further analyzing the flame snap-shot at $\hat{t}=2.17$ presented in Fig.~\ref{fig:ZH}, different quantities are plotted in Fig.~\ref{fig:Profiles} as a function of the scaled progress variable $Y$ (based on fuel mass fraction) and the scaled vertical distance $y/L$. The hat over the symbol indicates that the quantities have been scaled with the local maximum value. The strain rate $K_S$ decreases from the top wall to the bottom wall as can be expected from the applied boundary condition. It can also be noted that for $Y<0.3$ and $Y>0.85$, $K_S$ decreases drastically. The flame curvature can be computed as the divergence of the flame normal field as $\kappa= -\nabla \cdot \mathbf{n}$. It shows a negative value just at the top wall (as also observed in Fig.~\ref{fig:ZH}). $\kappa$ then increases close to $y/L \approx 0.125$ at $Y=0.9$. The flame displacement speed $S_{D,Y}$ as a field is computed using~\cite{jeroen1}

\begin{equation}
    S_{D,Y}=-   \frac{\nabla \cdot (\rho D_F \nabla Y_F)- \omega_F}{\rho_u  |\nabla Y_F|}.
\end{equation}

\noindent The above equation results from subtracting the flame kinematic equation from the progress variable transport equation. $D_F$ and $\nabla Y_F$ are the fuel diffusivity and progress variable gradient, respectively. The plot of $S_{D,Y}/S_L$ is dominated by the high value immediately close to the top wall but as observed previously, the flame does not burn there in a strong manner and the situation is similar to a flame cusp or flame tip for $Le < 1$ flames. In the immediate vicinity of this region, the flame segments have a speed between 2 and 1.5 times the laminar burning velocity corresponding to the region of high stretch (positive strain and negative stretch due to curvature). In the middle sections of the flame, $S_{D,Y} \leq S_L$ was found indicating weaker preferential diffusion effects.

\section{Dynamic stability conditions}  \addvspace{10pt}

\begin{figure*}[t!]
\centering
\includegraphics[width=190pt]{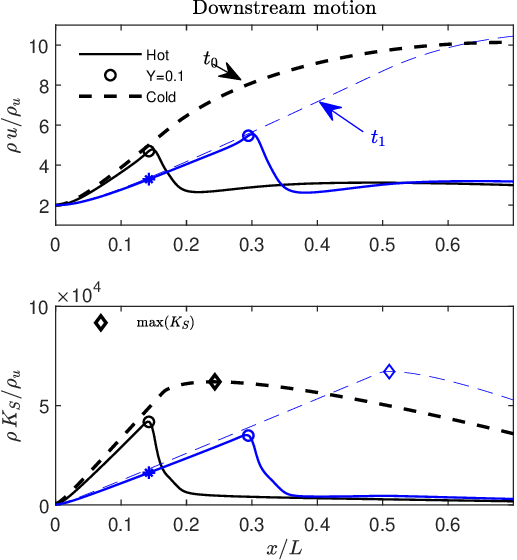}
\includegraphics[width=190pt]{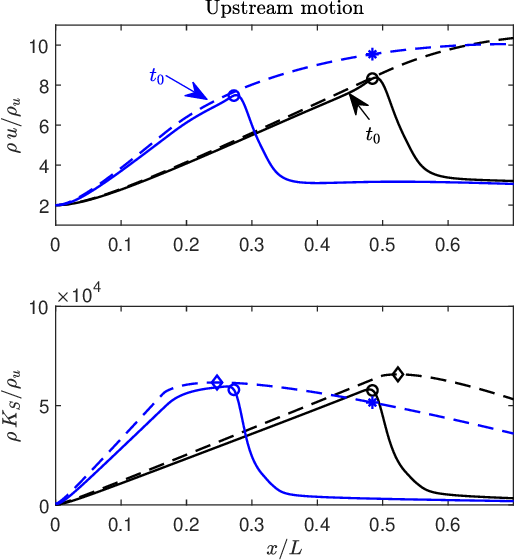}
\vspace{1pt}
\caption{Cold and hot flow (density averaged) profiles of x-velocity and strain rate at the location close to the top wall as a function of x-coordinate at two successive time-steps for the flame moving downstream and upstream.}
\label{fig:UK}
\vspace{-10pt}
\end{figure*}


In order to explain the conditions from the simulations that lead to the flame movement, the variation of the density weighted $x$-velocity $\rho u/\rho_\mathrm{u}$ in the $x$-direction is plotted in Fig.~\ref{fig:UK} for two time steps for a location close to the top wall. Results are included from the cold flow simulation (dashed) as well as the reactive flow simulation with the flame (solid) at the same flow-time in order to compare the gradients of the flow and strain rate. The location of the unburnt state ('u', identified by $Y=0.1$) is added along with the location where the strain $K_S$ shows its maximum. The results at $t_0$, which correspond to a time when the flame is closer to the inlet (for downstream motion and closer to the outlet for upstream motion), are plotted in black and results at a later time step $t_1$, with the flame moving further right, are plotted in blue. The downstream flame motion at $t_0$ is first analyzed assuming the applied boundary condition is frozen at $t_0$:

\begin{itemize}
    \item For the cold flow results, the velocity increases downstream of the flame location $\color{black}\circ$. The strain profile (cold) has a maximum ($\diamond$) downstream of the flame position ($\color{black}\circ$) and decreases beyond $\diamond$.
    \item This results in $du/dx>0$ and $dK_S/dx>0$ up to the $\diamond$ point. Beyond the $\diamond$ point, $dK_S/dx<0$ while $du/dx>0$ prohibiting the flame motion further downstream as the flame cannot increase its speed matching the increase in velocity.
    \item Thus, the condition in Eq.~\eqref{eq:dynstab} can be extended for the movement of the flame location downstream assuming the boundary condition is frozen, as
\end{itemize}

\begin{equation}
    \frac{\partial u}{\partial |n|} \approx \frac{\partial S_{D,Y}}{\partial K_S} \frac{\partial K_S}{\partial |n|} \,, \label{eq:DN}
\end{equation}

\noindent where the $x$-direction has been replaced by $n$, the flame normal direction. The above condition will allow the flame to move towards the right from $ \color{black} \circ$ up to $\diamond$ in a quasi-steady manner. 

With the time-varying boundary condition, the movement of the flame from $ \color{black} \circ$ at $t_0$ to $ \color{blue} \circ$ at $t_1$ can be analyzed as

\begin{itemize}
    \item The local velocity and the strain rate at $t_1$ at the old flame location is marked with a $\color{blue} \Asterisk$ symbol. It can be observed that $u$ and $K_S$ both decrease at the old location ($\color{blue} \Asterisk$) as a function of time (i.e.~from $t_0$ to $t_1$) due to the applied boundary condition. 
    \item As the velocity at $t_1$ at the old flame location ($\color{blue} \Asterisk$) is reduced, one could expect the flame to move upstream resulting in flashback due to the flame speed exceeding the local velocity. However, this is not the case.
    \item The decreasing strain at the old flame location ($\color{blue} \Asterisk$, marked on the $t_1$ plots), prohibits the flame from moving upstream. If the strain remained constant or increased at the $\color{blue} \Asterisk$ point, this could cause the flame to move upstream since the flame speed would no longer balance with the decreasing velocity at $\color{blue} \Asterisk$.
    \item Since the condition from Eq.~\eqref{eq:DN} remains valid spatially at $t_1$, this causes the flame to move downstream such that the flame moves from $\color{blue} \Asterisk$ towards $\color{blue} \diamond$, eventually stabilizing somewhat in between at $\color{blue} \circ$. 
    \item Given that Eq.~\eqref{eq:DN} is valid, the temporal dynamic stability condition at the old/default flame location that results in the flame moving downstream can be written as
\end{itemize}

\begin{equation}
    \frac{d K_s}{d t} < 0 \: \text{and} \: \frac{d u}{d t} < 0. 
\end{equation}

\noindent For the upstream moving flame, the same dynamic spatial stability condition in Eq.~\eqref{eq:DN} is valid as can be inferred from the right side plots in Fig.~\ref{fig:UK}. The temporal dynamic stability condition, however, is given as

\begin{equation}
    \frac{d K_s}{d t} < 0 \: \text{and} \: \frac{d u}{d t} > 0, 
\end{equation}

\noindent where the velocity at the old/default location ($\color{blue} \Asterisk$, marked for the plots of the new time step $t_1$) now increases with time while the strain still decreases with time. This causes the flame to move to a position on the left where lower velocity is present at a lower strain. Together with the spatial condition of Eq.~\eqref{eq:DN}, the temporal dynamic stability conditions cause the flame to move downstream and upstream in the channel in a periodic manner. 


\section{A note on practical outcomes from this study}

In this section, we present some of the possible practical outcomes of this study on the design of non-conventional burners/anchoring devices for hydrogen flames. As our model employed artificial boundary conditions, it is of importance here to state that the dynamic and period movement of the hydrogen flame could be utilized in a practical configuration. Insights from this work could also enable the design of burners which offer anchoring locations preferred by the hydrogen flames that are different from the conventional hydrocarbon flames. Currently, we present the following guidelines for the design of such devices in which hydrogen flames could be stabilized at high strain and high velocity anchoring locations:

\begin{itemize}
    \item Design of surfaces on which the shear stress could be controlled by, for example, having different surface roughness/structures which lead to a variation of local strain and velocity. In this way, by designing the surface of the channel walls, an experimental setup could be derived which reproduces the effects observed here.
    \item Results from this study could help in modifying the multi-slit geometry by varying the sharpness of the burner edge in multi-slit, bluff body type burners, adding extra steps etc. to generate region of higher strain and velocity.
    \item Another design could involve a geometrical element attached to a piston moving forward and backward along a section of the combustion chamber as shown in Fig.~\ref{fig:piston}. Such a moving element will disturb the flow thus generating locally high strain and result in high velocity due to motion of the element with respect to the inlet conditions.
 
 \begin{figure}
  \centering
{\includegraphics[width=\columnwidth]{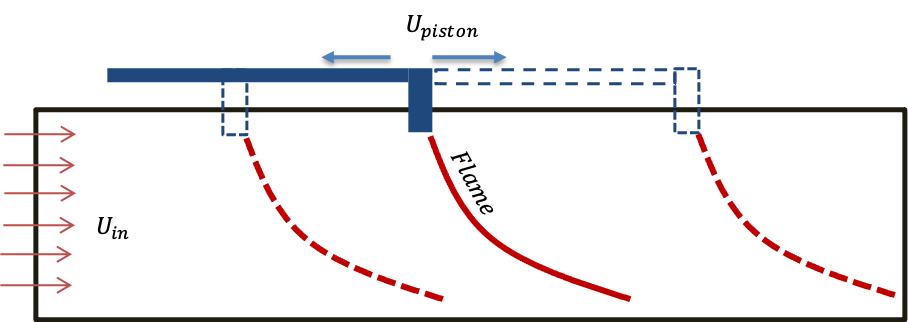}}
      \caption{Illustration of a geometrical element attached to a piston moving downstream and upstream of a reference location. The geometrical shape of the element could induce flow strain while the movement of the piston can increase/decrease the relative flow speed.}
      \label{fig:piston}
\vspace{-8pt}
\end{figure}

\end{itemize}

\section{Conclusion} \addvspace{10pt}
In this study, we have carried out an investigation into the movement of a hydrogen flame towards a preferred anchoring location of high strain and high velocity. For this purpose, the flame in a channel configuration has been modified by introducing a periodically moving high shear stress point at a wall. Such a point induces strain and velocity which enable the hydrogen flame to anchor near it in a dynamic manner and if the speed at which this point moves is kept below the laminar burning velocity, a periodic movement of the flame from the inlet towards the outlet and back towards the inlet was established. Dynamic stability conditions as a function of space and time have also been analyzed which show that if the flame speed gradient in the normal direction is equal to the velocity gradient, the flame could move towards the region of higher strain. Temporally, if the strain and the velocity at the default flame location increase/decrease, this could result in flame moving upstream/downstream given that the spatial dynamic stability condition for flame movement is valid. The novel modification of the flame in a channel configuration in this study allows for gaining further insights into the flame-flow interaction especially in the context of hydrogen flame stabilization.


\section*{Acknowledgements}
\label{Acknowledgments}
The research leading to these results has received funding from the European Union’s Horizon 2020 research and innovation program under the Center of Excellence in Combustion (CoEC) project, grant agreement No 952181.


\bibliography{publication.bib}

\begin{thebibliography}{51}
\providecommand{\natexlab}[1]{#1}
\providecommand{\url}[1]{\texttt{#1}}
\expandafter\ifx\csname urlstyle\endcsname\relax
  \providecommand{\doi}[1]{doi: #1}\else
  \providecommand{\doi}{doi: \begingroup \urlstyle{rm}\Url}\fi

\bibitem[Law(1989)]{Law1}
C.~K. Law.
\newblock {Dynamics of stretched flames}.
\newblock \emph{Symp. (Int.) Combust.}, 22:\penalty0 1381--1402, 1989.

\bibitem[Poinsot and Veynante(2005)]{Poinsotb}
T.~Poinsot and D.~Veynante.
\newblock \emph{Theoretical and Numerical Combustion}.
\newblock Edwards, 2005.

\bibitem[Vance et~al.(2020)Vance, Shoshin, van Oijen, and Goey]{vance2}
F.~H. Vance, Y.~Shoshin, J.~A. van Oijen, and L.~P.~H. Goey.
\newblock {The effect of thermal diffusion on stabilization of premixed
  flames}.
\newblock \emph{Combust. Flame}, 216:\penalty0 45--57, 2020.

\bibitem[Jiménez et~al.(2018)Jiménez, Michaels, and Ghoniem]{cimenez}
C.~Jiménez, D.~Michaels, and A.~F. Ghoniem.
\newblock {Stabilization of ultra-lean hydrogen enriched inverted flames behind
  a bluff–body and the phenomenon of anomalous blow–off}.
\newblock \emph{Combust. Flame}, 191:\penalty0 86--98, 2018.

\bibitem[Michaels et~al.(2017)Michaels, Shanbhogue, and Ghoniem]{Dan1}
D.~Michaels, S.~J. Shanbhogue, and A.~F. Ghoniem.
\newblock {The impact of reactants composition and temperature on the flow
  structure in a wake stabilized laminar lean premixed CH4/H2/air flames;
  mechanism and scaling}.
\newblock \emph{Combust. Flame}, 176:\penalty0 151--161, 2017.

\bibitem[Vance et~al.(2021{\natexlab{a}})Vance, Shoshin, van Oijen, and
  Goey]{vancesc}
F.~H. Vance, Y.~Shoshin, J.~A. van Oijen, and L.~P.~H. Goey.
\newblock A physical relationship between consumption and displacement speed
  for premixed flames with finite thickness.
\newblock \emph{Proc. Combust. Inst.}, 38, 2021{\natexlab{a}}.

\bibitem[van Oijen et~al.(2016)van Oijen, Donini, Bastiaans, ten~thije
  Boonkkamp, and de~Goey]{state}
J.~A. van Oijen, A.~Donini, R.~J.~M. Bastiaans, J.~H.~M. ten~thije Boonkkamp,
  and L.~P.~H. de~Goey.
\newblock {State-of-the-art in premixed combustion modeling using flamelet
  generated manifolds}.
\newblock \emph{Prog. Energ. Combust.}, 57:\penalty0 30--74, 2016.

\bibitem[Vance et~al.(2022{\natexlab{a}})Vance, de~Goey, and van
  Oijen]{vancefb}
F.~H. Vance, L.~P.~H. de~Goey, and J.~A. van Oijen.
\newblock {Development of a flashback correlation for burner-stabilized
  hydrogen-air premixed flames}.
\newblock \emph{Combust. Flame}, 235, 2022{\natexlab{a}}.

\bibitem[Konnov(2019)]{Kon}
A.~A. Konnov.
\newblock {Yet another kinetic mechanism for hydrogen combustion}.
\newblock \emph{Combust. Flame}, 203:\penalty0 14--22, 2019.

\bibitem[Kazakov and Frenklach()]{DRM}
A.~Kazakov and M.~Frenklach.
\newblock {Reduced Reaction Sets based on GRI-Mech 1.2}.
\newblock URL \url{http://combustion.berkeley.edu/drm/}.

\bibitem[Vance et~al.(2022{\natexlab{b}})Vance, Shoshin, de~Goey, and van
  Oijen]{vancenrz}
F.~H. Vance, Y.~Shoshin, L.~P.~H. de~Goey, and J.~A. van Oijen.
\newblock {Quantifying the impact of heat loss, stretch and preferential
  diffusion effects to the anchoring of bluff body stabilized premixed flames}.
\newblock \emph{Combust. Flame}, 237, 2022{\natexlab{b}}.

\bibitem[Shoshin et~al.(2013)Shoshin, Bastiaans, and de~Goey]{yuri}
Y.~Shoshin, R.~Bastiaans, and L.~P.~H. de~Goey.
\newblock {Anomalous blow-off behavior of laminar inverted flames of ultra-lean
  hydrogen–methane–air mixtures}.
\newblock \emph{Combust. Flame}, 160:\penalty0 565--576, 2013.

\bibitem[Lewis and von Elbe(1943)]{Elbe1}
B.~Lewis and G.~von Elbe.
\newblock {Stability and structure of Burner Flames}.
\newblock \emph{J. Chem. Phys.}, 75, 1943.

\bibitem[Kedia and Ghoniem(2015)]{Kedia2}
K.~S. Kedia and A.~F. Ghoniem.
\newblock {The blow-off mechanism of a bluff-body stabilized laminar premixed
  flame}.
\newblock \emph{Combust. Flame}, 162:\penalty0 1304--1315, 2015.

\bibitem[Kawamura et~al.(1982)Kawamura, Asato, and Mazaki]{kawa1}
T.~Kawamura, K.~Asato, and T.~Mazaki.
\newblock {Reexamination of the blowoff mechanism of premixed flames –
  inverted flames}.
\newblock \emph{Combust. Flame}, 45:\penalty0 225--233, 1982.

\bibitem[Giannakopoulos et~al.(2019)Giannakopoulos, Frouzakis, Mohan,
  Tomboulides, and Matalon]{mat2}
G.~K. Giannakopoulos, C.~E. Frouzakis, S.~Mohan, A.~G. Tomboulides, and
  M.~Matalon.
\newblock {Consumption and displacement speeds of stretched premixed flames -
  Theory and simulations}.
\newblock \emph{Combust. Flame}, 208:\penalty0 164--181, 2019.

\bibitem[Klukas et~al.(2020)Klukas, Giglmaier, Adams, Sieber, Schimek, and
  Paschereit]{KLUKAS202032547}
S.~Klukas, M.~Giglmaier, N.~A. Adams, M.~Sieber, S.~Schimek, and C.~O.
  Paschereit.
\newblock Anchoring of turbulent premixed hydrogen/air flames at externally
  heated walls.
\newblock \emph{Int. J. Hydrogen Energ.}, 45:\penalty0 32547--32561, 2020.

\bibitem[Golovastov et~al.(2021)Golovastov, Bivol, and
  Golub]{GOLOVASTOV20212783}
S.~V. Golovastov, G.~Y. Bivol, and V.~V. Golub.
\newblock Influence of porous walls on flame front perturbations in
  hydrogen-air mixtures.
\newblock \emph{Int. J. Hydrogen Energ.}, 46:\penalty0 2783--2795, 2021.

\bibitem[Zhang et~al.(2020)Zhang, Chang, Wang, and Huang]{ZHANG20201072}
M.~Zhang, M.~Chang, J.~Wang, and Z.~Huang.
\newblock Flame dynamics analysis of highly hydrogen-enrichment premixed
  turbulent combustion.
\newblock \emph{Int. J. Hydrogen Energ.}, 45:\penalty0 1072--1083, 2020.

\bibitem[Elshimy et~al.(2020)Elshimy, Ibrahim, and
  Malalasekera]{ELSHIMY202014979}
M.~Elshimy, S.~Ibrahim, and W.~Malalasekera.
\newblock Numerical studies of premixed hydrogen/air flames in a small-scale
  combustion chamber with varied area blockage ratio.
\newblock \emph{Int. J. Hydrogen Energ.}, 45:\penalty0 14979--14990, 2020.

\bibitem[Wu et~al.(2021)Wu, Zheng, Dong, Zhang, and Ding]{WU202133601}
H.~Wu, J.~Zheng, X.~Dong, S.~Zhang, and Y.~Ding.
\newblock Investigations on the cellular instabilities of expanding
  hydrogen/methanol spherical flame.
\newblock \emph{Int. J. Hydrogen Energ.}, 46:\penalty0 33601--33615, 2021.

\bibitem[Zou et~al.(2020)Zou, Deng, Kang, and Wang]{ZOU202020436}
P.~Zou, Y.~Deng, X.~Kang, and J.~Wang.
\newblock A numerical study on premixed hydrogen/air flames in a narrow channel
  with thermally orthotropic walls.
\newblock \emph{Int. J. Hydrogen Energ.}, 45:\penalty0 20436--20448, 2020.

\bibitem[Marragou et~al.(2022)Marragou, Magnes, Poinsot, Selle, and
  Schuller]{MARRAGOU202219275}
S.~Marragou, H.~Magnes, T.~Poinsot, L.~Selle, and T.~Schuller.
\newblock Stabilization regimes and pollutant emissions from a dual fuel ch4/h2
  and dual swirl low nox burner.
\newblock \emph{Int. J. Hydrogen Energ.}, 47:\penalty0 19275--19288, 2022.

\bibitem[Vance et~al.(2021{\natexlab{b}})Vance, Shoshin, de~Goey, and van
  Oijen]{en14071977}
F.~H. Vance, Y.~Shoshin, L.~P.~H. de~Goey, and J.~A. van Oijen.
\newblock Flame stabilization and blow-off of ultra-lean $\ce{H2}$-air premixed
  flames.
\newblock \emph{Energies}, 14, 2021{\natexlab{b}}.

\bibitem[Lee et~al.(2015)Lee, Yoo, and Im]{LEE20152602}
B.~J. Lee, C.~S. Yoo, and H.~G. Im.
\newblock Dynamics of bluff-body-stabilized premixed hydrogen/air flames in a
  narrow channel.
\newblock \emph{Combust. Flame}, 162:\penalty0 2602--2609, 2015.

\bibitem[Jiménez et~al.(2015)Jiménez, Fernández-Galisteo, and
  Kurdyumov]{JIMENEZ201512541}
C.~Jiménez, D.~Fernández-Galisteo, and V.~N. Kurdyumov.
\newblock Dns study of the propagation and flashback conditions of lean
  hydrogen-air flames in narrow channels: Symmetric and non-symmetric
  solutions.
\newblock \emph{Int. J. Hydrogen Energ.}, 40:\penalty0 12541--12549, 2015.

\bibitem[Reichel and Paschereit(2017)]{REICHEL20174518}
T.~G. Reichel and C.~O. Paschereit.
\newblock Interaction mechanisms of fuel momentum with flashback limits in
  lean-premixed combustion of hydrogen.
\newblock \emph{Int. J. Hydrogen Energ.}, 42:\penalty0 4518--4529, 2017.

\bibitem[Endres and Sattelmayer(2018)]{ENDRES2018151}
A.~Endres and T.~Sattelmayer.
\newblock Large eddy simulation of confined turbulent boundary layer flashback
  of premixed hydrogen-air flames.
\newblock \emph{Int. J. Heat Fluid FL.}, 72:\penalty0 151--160, 2018.

\bibitem[Goldmann and Dinkelacker(2022)]{GOLDMANN2022111927}
A.~Goldmann and F.~Dinkelacker.
\newblock Investigation of boundary layer flashback for non-swirling premixed
  hydrogen/ammonia/nitrogen/oxygen/air flames.
\newblock \emph{Combust. Flame}, 238:\penalty0 111927, 2022.

\bibitem[Wan and Zhao(2020)]{WAN2020358}
J.~Wan and H.~Zhao.
\newblock Blow-off mechanism of a holder-stabilized laminar premixed flame in a
  preheated mesoscale combustor.
\newblock \emph{Combust. Flame}, 220:\penalty0 358--367, 2020.

\bibitem[Chen(2009)]{CHEN20096558}
Z.~Chen.
\newblock Effects of hydrogen addition on the propagation of spherical
  methane/air flames: A computational study.
\newblock \emph{Int. J. Hydrogen Energ.}, 34:\penalty0 6558--6567, 2009.

\bibitem[Shi et~al.(2016)Shi, Chen, and Chen]{SHI2016394}
X.~Shi, J.-Y. Chen, and Z.~Chen.
\newblock Numerical study of laminar flame speed of fuel-stratified
  hydrogen/air flames.
\newblock \emph{Combust. Flame}, 163:\penalty0 394--405, 2016.

\bibitem[Zhang and Chen(2013)]{doi:10.1080/13647830.2013.792393}
H.~Zhang and Z.~Chen.
\newblock Effects of heat conduction and radical quenching on premixed
  stagnation flame stabilised by a wall.
\newblock \emph{Combust. Theor. Model.}, 17:\penalty0 682--706, 2013.

\bibitem[Zhang et~al.(2017)Zhang, Zirwes, Habisreuther, and
  Bockhorn]{ZHANG2017170}
F.~Zhang, T.~Zirwes, P.~Habisreuther, and H.~Bockhorn.
\newblock Effect of unsteady stretching on the flame local dynamics.
\newblock \emph{Combust. Flame}, 175:\penalty0 170--179, 2017.

\bibitem[Zhang et~al.(2021)Zhang, Zirwes, Häber, Bockhorn, Trimis, and
  Suntz]{ZHANG20211955}
F.~Zhang, T.~Zirwes, T.~Häber, H.~Bockhorn, D.~Trimis, and R.~Suntz.
\newblock Near wall dynamics of premixed flames.
\newblock \emph{Proc. Combust. Inst.}, 38:\penalty0 1955--1964, 2021.

\bibitem[Gerlinger et~al.(2003)Gerlinger, Schneider, Fröhlich, and
  Bockhorn]{GERLINGER2003247}
W.~Gerlinger, K.~Schneider, J.~Fröhlich, and H.~Bockhorn.
\newblock Numerical simulations on the stability of spherical flame structures.
\newblock \emph{Combust. Flame}, 132:\penalty0 247--271, 2003.

\bibitem[Maruta(2011)]{maruta}
K.~Maruta.
\newblock {Micro and mesoscale combustion}.
\newblock \emph{Proc. Combust. Inst.}, 33:\penalty0 125--150, 2011.

\bibitem[Dejoan and Kurdyumov(2019)]{kurd6}
A.~Dejoan and V.~N. Kurdyumov.
\newblock {Thermal expansion effect on the propagation of premixed flames in
  narrow channels of circular cross-section: Multiplicity of solutions,
  axisymmetry and non-axisymmetry}.
\newblock \emph{Proc. Combust. Inst.}, 37:\penalty0 1927--1935, 2019.

\bibitem[Veiga-López et~al.(2020)Veiga-López, Martínez-Ruiz, Kuznetsov, and
  Sánchez-Sanz]{spain}
F.~Veiga-López, D.~Martínez-Ruiz, M.~Kuznetsov, and M.~Sánchez-Sanz.
\newblock {Thermoacoustic analysis of lean premixed hydrogen flames in narrow
  vertical channels}.
\newblock \emph{Fuel}, 278, 2020.

\bibitem[Kurdyumov and Matalon(2016)]{kurd}
V.~N. Kurdyumov and M.~Matalon.
\newblock {Effects of gas compressibility on the dynamics of premixed flames in
  long narrow adiabatic channels}.
\newblock \emph{Combust. Theory Model.}, 20:\penalty0 1046 -- 1067, 2016.

\bibitem[Ronney(2003)]{ronney}
P.~D. Ronney.
\newblock {Analysis of non-adiabatic heat-recirculating combustors}.
\newblock \emph{Combust. Flame}, 135:\penalty0 421--439, 2003.

\bibitem[Bioche et~al.(2019)Bioche, Ribert, and Vervisch]{vervisch}
K.~Bioche, G.~Ribert, and L.~Vervisch.
\newblock {Simulating upstream flame propagation in a narrow channel after wall
  preheating: Flame analysis and chemistry reduction strategy}.
\newblock \emph{Combust. Flame}, 200:\penalty0 219--231, 2019.

\bibitem[Schlup and Blanquart(2018)]{model}
J.~Schlup and G.~Blanquart.
\newblock {A reduced thermal diffusion model for H and H2}.
\newblock \emph{Combust. Flame}, 191:\penalty0 1--8, 2018.

\bibitem[Barlow et~al.(2001)Barlow, Karpetis, Frank, and Chen]{rad}
R.~S. Barlow, A.~N. Karpetis, J.~H. Frank, and J.~Y. Chen.
\newblock {Scalar profiles and NO formation in laminar opposed-flow partially
  premixed methane/air flames}.
\newblock \emph{Combust. Flame}, 127:\penalty0 2102--2118, 2001.

\bibitem[Li et~al.(2003)Li, Zhao, Kazakov, and Dryer]{Li}
J.~Li, Z.~Zhao, A.~Kazakov, and F.~L. Dryer.
\newblock {An Updated Comprehensive Kinetic Model for H2 Combustion}.
\newblock \emph{Fall Technical Meeting of the Eastern States Section of the
  Combustion Institute, Penn State University, University Park, PA}, 200, 2003.

\bibitem[Ans(2020)]{Ansys}
{Ansys® Fluent, Release 20.2}.
\newblock 2020.

\bibitem[che(2021)]{chem1d}
{CHEM1D. A one dimensional laminar flame code. Eindhoven University of
  Technology}.
\newblock 2021.

\bibitem[Vance et~al.(2019)Vance, Shoshin, van Oijen, and Goey]{vance1}
F.~H. Vance, Y.~Shoshin, J.~A. van Oijen, and L.~P.~H. Goey.
\newblock {Effect of Lewis number on premixed laminar lean-limit flames
  stabilized on a bluff body}.
\newblock \emph{Proc. Combust. Inst.}, 37:\penalty0 1663--1672, 2019.

\bibitem[Burali et~al.(2016)Burali, Lapointe, Bobbitt, Blanquart, and
  Xuan]{ConstLE}
N.~Burali, S.~Lapointe, B.~Bobbitt, G.~Blanquart, and Y.~Xuan.
\newblock {Assessment of the constant non-unity Lewis number assumption in
  chemically-reacting flows}.
\newblock \emph{Combust. Theory Model.}, 20:\penalty0 632--657, 2016.

\bibitem[van Oijen(2002)]{oijen1}
J.~A. van Oijen.
\newblock \emph{{Flamelet-generated manifolds : development and application to
  premixed laminar flames}}.
\newblock PhD thesis, 2002.

\bibitem[van Oijen et~al.(2005)van Oijen, Groot, Bastiaans, and
  de~Goey]{jeroen1}
J.~A. van Oijen, G.~R.~A. Groot, R.~J.~M. Bastiaans, and L.~P.~H. de~Goey.
\newblock {A flamelet analysis of the burning velocity of premixed turbulent
  expanding flames}.
\newblock \emph{Proc. Combust. Inst.}, 30, 2005.

\end{thebibliography}
\bibliographystyle{unsrtnat_mod}

\end{document}